\documentclass{article}
\usepackage[T1]{fontenc} 
\usepackage[utf8]{inputenc} 
\usepackage{ismir,amsmath,cite,url}
\usepackage{graphicx}
\usepackage[table,xcdraw]{xcolor}
\usepackage{color}
\usepackage{dblfloatfix}
\usepackage{subcaption}

\usepackage{lineno}
\usepackage{tabularx}
\newenvironment{conditions*}
  {\par\vspace{\abovedisplayskip}\noindent
   \tabularx{\columnwidth}{>{$}l<{$} @{${}={}$} >{\raggedright\arraybackslash}X}}
  {\endtabularx\par\vspace{\belowdisplayskip}}
 
\newcommand{\customfootnotetext}[2]{{
  \renewcommand{\thefootnote}{#1}
  \footnotetext[0]{#2}}}
 
\twocolumn
  
\title{Tag2Risk: Harnessing Social Music Tags for Characterizing Depression Risk}

\multauthor
{Aayush Surana$^{*1}$ \hspace{1cm} Yash Goyal$^{*1}$ \hspace{1cm} Manish Shrivastava$^1$} { \bfseries{Suvi Saarikallio$^2$ \hspace{1cm} Vinoo Alluri$^1$}\\
 $^1$ International Institute of Information Technology, Hyderabad, India \\$^2$ Department of Music, Art and Culture Studies, University of Jyväskylä, Finland\\
{\tt\small \{aayush.surana, yash.goyal\}@research.iiit.ac.in, suvi.saarikallio@jyu.fi}\\
{\tt\small \{m.shrivastava, vinoo.alluri\}@iiit.ac.in}
}
\def\authorname{A. Surana, Y. Goyal, M. Shrivastava, S. Saarikallio, and V. Alluri}

\begin{document}

\maketitle
\begin{abstract}
Musical preferences have been considered a mirror of the self. In this age of Big Data, online music streaming services allow us to capture ecologically valid music listening behavior and provide a rich source of information to identify several user-specific aspects. Studies have shown musical engagement to be an indirect representation of internal states including internalized symptomatology and depression. The current study aims at unearthing patterns and trends in the individuals at risk for depression as it manifests in naturally occurring music listening behavior. Mental well-being scores, musical engagement measures, and listening histories of Last.fm users (N=541) were acquired. Social tags associated with each listener's most popular tracks were analyzed to unearth the mood/emotions and genres associated with the users. Results revealed that social tags prevalent in the users at risk for depression were predominantly related to emotions depicting \textit{Sadness} associated with genre tags representing \textit{neo-psychedelic-, avant garde-, dream-pop}. This study will open up avenues for an MIR-based approach to characterizing and predicting risk for depression which can be helpful in early detection and additionally provide bases for designing music recommendations accordingly.

\customfootnotetext{$*$}{ - Joint first authors with equal contribution.}

\end{abstract}
\begin{figure*}[h]\label{methodology}
  \centering
  \includegraphics[width=0.78\textwidth]{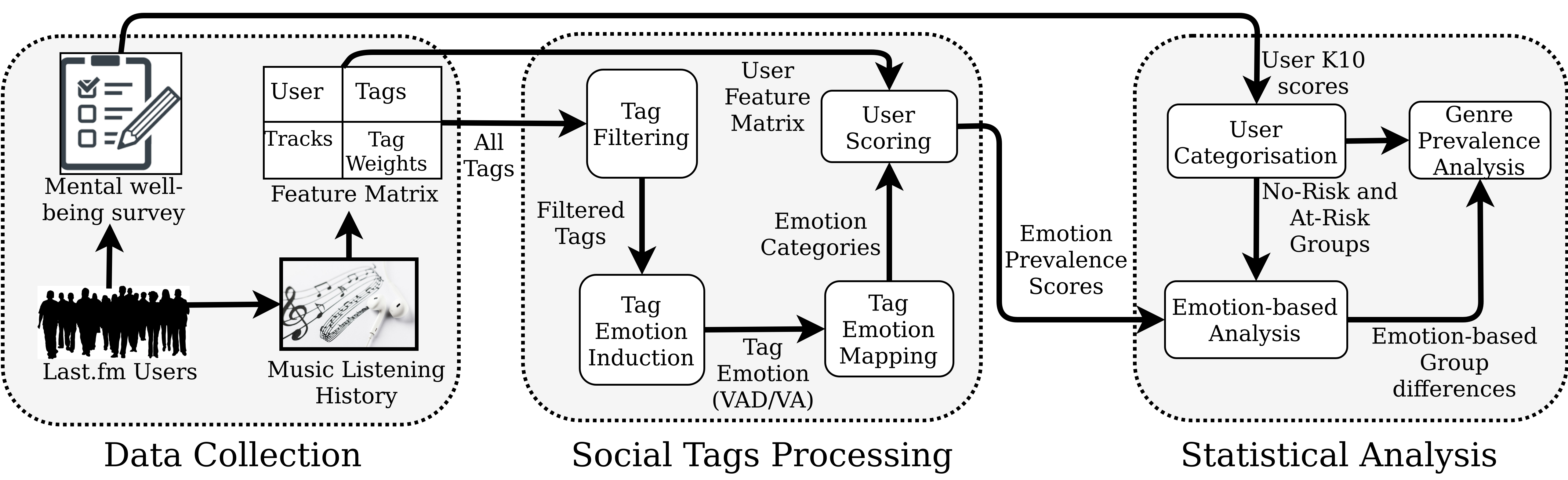}
  \caption{Methodology}
  \label{methodology}
\end{figure*}
\section{Introduction}\label{sec:introduction}
According to reports from the World Health Organization, an estimated 322 million people worldwide were affected from depression, the leading cause of disability \cite{world2017depression}. Recent times have witnessed a surge in studies on using social multimedia content, such as those from Facebook, Twitter, Instagram, to detect mental disorders including depression \cite{munmun,copper,munmun_14(2),munmun_16,reece}. Music plays a vital role in mental well-being by impacting moods, emotions and other affective states \cite{baltazar}. Musical preferences and habits have been associated with the individual’s need to satisfy and reinforce their psychological needs \cite{nave,qiu}. Empirical evidence exists linking musical engagement strategies to measures of ill-health including internalized symptomatology and depression \cite{litlink,rumin}. Also, increased emotional dependency on music during periods of depression has been reported \cite{mcferran}. Specifically, the Healthy-Unhealthy Music Scale (HUMS), a 13-item questionnaire was developed for assessing musical engagement strategies that identified maladaptive ways of using music. Such strategies are characterized by using music to avoid other people, resort to ruminative thinking and ending up feeling worse after music engagement. Such unhealthy musical engagement was found to correlate with  higher psychological distress and 
was indicative of depressive tendencies \cite{hums}. 
Furthermore, the high predictive power observed from the machine learning models in predicting risk for depression from HUMS further bolsters its efficacy as an indirect tool for assessing mental states \cite{ragarwal}. Research suggests that such musical engagement does not always lead to alleviating depressive symptoms \cite{stewart}. This indeed calls for developing intervention strategies that allow for altering music listening behavior to suit the individual’s state, traits, and general musical preferences which may lead to a positive outcome. Thus, it is of vital importance not only to identify individuals with depressive tendencies but also to unearth music listening habits of such individuals that will provide bases for designing music recommendations accordingly.

Past research studying the link between music listening habits and depression has been done using self-reported data and controlled listening experiments wherein participants may have wished to conform to social expectations, or their responses might be influenced by how they want other people to perceive them thereby resulting in demand characteristics \cite{greenberg2017social}. This has also been identified as a limitation by Nave et al. \cite{nave}, who have proposed collecting data in more ecologically valid settings, such as user listening histories from music streaming platforms which are a better reflection of the users’ true preferences and behaviours. To date no studies have looked at the link between active music listening and depression using the music listening histories of users which motivates us for this study.

In this age of big data, online music streaming platforms such as Last.fm, Spotify, and Apple Music provide access to millions of songs of varying genres and this has allowed for assessing users' features from their digital traces on music streaming platforms. To the best of our knowledge, Last.fm is the only platform that makes it possible to extract the listening history of users and other metadata describing their listening behavior using its own public API. Last.fm has been used extensively by researchers for various purposes such as music emotion classification, user behavior analysis, and social tag categorization \cite{saari,laurier}. Last.fm has an abundance of social tags that provide a wide range of information about the musical tracks including audio low- and high-level feature description, emotions and experiences evoked, genre, etc. These tags have been found to predict short-term user music preferences \cite{gupta} and in successfully predicting next played songs in the design of a recommendation system \cite{polignano}. Our aim is to identify the tags and their respective occurrences in the listening behavior of individuals at risk for depression, which makes Last.fm an apt choice for this study. The data was collected using an online survey comprising of Last.fm music listening histories, in addition to music engagement strategies (HUMS), and mental well-being scores of the participants. Specifically, each track in the data was semantically represented by the tags assigned to it. We leverage these representations of tags as social descriptors of music to uncover emotional experiences and concepts that are associated with users with risk for depression.

\subsection{Research Objectives and Hypotheses}
In this study we investigated whether people’s music listening history, in terms of social tags, could be used to predict a risk for depression.  Our research questions were:
\begin{itemize}
\item What are the social tags associated with music chosen by At-Risk users?
\item What emotions do these tags signify in the context of musically evoked emotions?
\item What genres are mostly associated with At-Risk users?
\item How well can we classify users as At-Risk given user-specific social tags?
\end{itemize}
We expected the social tags linked with At-Risk listeners to contain emotions with low arousal and low valence, being typical of depressive mood. Owing to the lack of research associating music genres and risk for depression\cite{stewart}, this part of the study was exploratory.

\section{Methodology}\label{sec:page_size}
The methodological approach and procedure of our study is illustrated in \figref{methodology}.
The steps of data collection, processing, and analysis are described below.
\subsection{Data Collection}
An online survey was designed wherein participants were asked to fill their Last.fm usernames and demographics followed by standard scales for assessing their mental well-being, musical engagement strategies and personality. Participants were solicited on the Last.fm groups of social media platforms like Reddit and Facebook. The inclusion criterion required being an active listener on Last.fm for at least a year prior to filling the survey. The survey form required the users’ consent to access their Last.fm music history.

\subsubsection{Participants}
A total of 541 individuals (Mean Age = 25.4, SD = 7.3) were recorded to be eligible and willing to participate in the study consisting of 444 males, 82 females and 15 others. Most of them belonged to the United States and the United Kingdom accounting for about 30\% and 10\% of the participants respectively. Every other country contributed to less than 5\% of the total participants.

\subsubsection{Measure of Well-Being, Musical Engagement, and Personality}
The Kessler’s Psychological Distress Scale (K10) questionnaire \cite{k10} was used to assess mental well-being. It is a measure of psychological distress, particularly assessing anxiety and depression symptoms. Individuals scoring 29 and above on K-10 are likely to be at severe risk for depression and hence, constitute the "At-Risk" group. Those scoring below 20 are labeled as the "No-Risk" group \cite{sakka} as they are likely to be well. There were 193 participants in the No-Risk group and 142 in the At-Risk group. The HUMS survey was administered to assess musical engagement strategies which resulted in two scores per participant, \textit{Healthy} and \textit{Unhealthy}. Personality information was obtained using the Mini-IPIP questionnaire \cite{topo} which results in scores for the Big Five traits of Personality namely \textit{Openness}, \textit{Conscientiousness}, \textit{Extraversion}, \textit{Agreeableness} and \textit{Neuroticism}. HUMS and personality data were collected in order to identify if specific personality traits engage more in \textit{Unhealthy} music listening and as additional measures to assess internal validity.

\subsubsection{Music Listening History}
Each participant’s music listening history was extracted using a publicly available API. The data included tracks, artists, and social tags associated with the tracks. For each participant, the top \textit{n} (n=500,200,100) tracks based on play-counts were extracted centered around the time \textit{t} (t = $\pm$ 3 months,2 months) they filled in the questionnaire. The reason for varying \textit{n} and \textit{t} was to find converging evidence in music listening behavior in order to make our results more robust. For each track, the top 50 social tags based on tag weight (number of times the tag has been assigned to the track) were chosen for subsequent analysis.

\subsection{Social Tags Processing}
\subsubsection{Tag Filtering}
Music-related social tags are known to be descriptors of genre, perceived emotion, artist and album amongst others. It is therefore important to filter them to organize them according to some structure and interpretable dimensions for the task at hand. The purpose of this preprocessing step was to retrieve tags that could be mapped onto a semantic space representing music-evoked emotions. To this end, we used four filtering stages: first, include lower-casing, removal of punctuation and stop-words, spell-checking and checking for the existence of tag words in the English corpus; second, retain tags that are most frequently assigned adverbs or adjectives via POS (Part Of Speech) tagging since POS tags representing nouns and pronouns do not have emotion relevance in this context; third, remove tags containing 2 or more words to avoid valence shifters \cite{valence_shifters} and sentence-like descriptions from our Last.fm corpus; fourth, manually filter them by discarding tags without any mood/emotion associations.


\subsubsection{Tag Emotion Induction}\label{induction}
To project the tags onto an emotion space, we used dimensional models that represent the emotions. Multiple research studies have shown the usefulness of both two-dimensional and three-dimensional models to represent emotions \cite{dimension1, eurola, dimension2}. We therefore used both these models for further analysis in order to check for trends and the effect of the third dimension when dealing with emotions. 

The first model is one of the most popular dimensional models, the Russell's Circumplex Model of Affect \cite{VAmodel}, where an emotion is a point in a two-dimensional continuous space representing \textit{Valence} and \textit{Arousal} (VA). \textit{Valence} reflects pleasantness and \textit{Arousal} describes the energy content of the emotion. The second model is an extension of the Russell's model with an added \textit{Dominance} dimension (VAD), which represents control of the emotional state. The VAD model has been a popular framework used to construct emotion lexicons in the field of Natural Language Processing. The projection in the VAD space is based on semantic similarity and has been largely used to obtain affective ratings for large corpora of English words \cite{wei} \cite{vad}.
Another common emotion model is the VAT model wherein the third dimension represents \textit{Tension} (VAT) and has been used in retrieving mood information from Last.fm tags \cite{saari}. However, Saari et. al.'s \cite{saari} approach was based on tag co-occurrence rather than semantic similarity. Moreover, a subsequent study by the same authors reported a positive correlation (r=0.85) between \textit{tension} and \textit{dominance} \cite{paasi_2}. Also, multiple studies have supported the use of the VAD space for analysing emotions in the context of music \cite{musicvad, musicvad2}. We therefore have chosen the VAD framework for the purpose of our study. Since VA dimensions alone were found to sufficiently capture musical emotions \cite{eurola}, we also repeat our analysis based on the VA model to observe the effect of the omitted \textit{Dominance} dimension.


The tags were projected onto the VAD space using a word-emotion induction model introduced by Buechel and Hahn \cite{wei}. We used the FastText embeddings of the tags as input to a 3-layer multi-layer perceptron that produced VAD values ranging from 1 to 9 on either of the dimensions. FastText has shown better accuracy for word-emotion induction \cite{wei} when compared to other commonly used models like Word2vec and GloVe. Moreover, FastText embeddings incorporate sub-word character n-grams that enable the handling of out-of-vocabulary words. This results in a large advantage over the other models \cite{fasttext}. In addition, FastText works well with rarely occurring words because their character n-grams are still shared with other words. This made it a suitable choice since some of the user-assigned tags may be infrequent or absent in the training corpus of the embedding model. We used the same approach to project the tags onto the VA space by changing the number of nodes in the output layer from 3 to 2. 

Both the models were trained using the EN+ dataset which contains \textit{valence}, \textit{arousal} and \textit{dominance} ratings (on a 9-point scale) for a majority of well-known English words \cite{Warriner2013}. This module resulted in an n-dimensional vector (n=3 for VAD, n=2 for VA) for each tag. The remainder of the pipeline describes the 3-dimensional VAD vector processing. The same procedure is repeated for the VA scores.

\subsubsection{Tag Emotion Mapping}\label{clustering}
The social tags were grouped into broader emotion categories. These categories consisted of 9 first-order factors of Geneva Emotional Music Scale (GEMS) \cite{gems}. These were \textit{Wonder, Transcendence, Nostalgia, Tenderness, Peacefulness, Power, Joyful activation, Tension and sadness}. Table 1 in the supplementary material displays the factor loadings for these first-order factors of GEMS.
GEMS contains 40 emotion terms that were consistently chosen to describe musically evoked emotive states across a wide range of music genres. These were subsequently grouped to provide a taxonomy for music-evoked emotions. This scale has outperformed other discrete and dimensional emotion models in accounting for music-evoked emotions \cite{gemsvsVAD}. In order to project these 9 emotion categories onto the VAD space, we first obtained the VAD values for the 40 emotion terms. Next, the VAD values were weighted and summed according to the weights provided in the original GEMS study to finally obtain VAD values for each of the emotion categories. Figures 1 \& 2 in supplementary material display the projections of these emotion categories onto the VAD and VA spaces. Each of the tags are then assigned the emotion category based on the proximity in the VAD space as evaluated by the euclidean distance.  

\subsection{User-Specific Emotion Prevalence Score}\label{scoring}
After every user’s tags had been mapped onto the 9 emotion categories,we calculated an \textit{Emotion Prevalence Score} $S_{u,c}$ for every user. This represents the presence of tags belonging to that particular emotion category in the user's listening history.
\begin{equation}\label{userscore}
S_{u,c}=\frac{\sum_{j\epsilon{V\textsubscript{tr}}}\left(N_{j,c}\times tr\textsubscript{u,j}\right)}{\sum_{i\epsilon{T\textsubscript{u}}} tr \textsubscript{u,i}}
\end{equation}
where \begin{equation}\label{n_tagweight}
N_{j,c}=\sum_{k\epsilon{Tags\textsubscript{c}}}\frac{tw\textsubscript{j,k}}{\sum_{l\epsilon{V\textsubscript{tg}}}tw\textsubscript{j,l}}
\end{equation}
$c$ : emotion category\\
$N_{j,c}$ : the association of track $j$ with $c$ \\ $T_u$ : all tracks for user $u$\\ $V_{tg}$ : all tags obtained after tag filtering\\ $V_{tr}$ : all tracks having at least one tag from $V_{tg}$\\ $tr_{u,i}$ : playcount of track $i$ for user $u$\\ $tw_{j,k}$ : tag weight of tag $k$ for track $j$ \\ $Tags_c$ : all tags in $V_{tg}$ which belong to $c$ \\

Since the objective of this work was to identify which of the 9 categories are most characteristic of At-Risk individuals when compared to No-Risk individuals, we performed group-level statistical tests of difference as described in the following section.
\subsection{Emotion-based Analysis : Group Differences and Bootstrapping}
For each emotion category, we performed a two-tailed Mann-Whitney U (MWU) Test on the \textit{Emotion Prevalence Scores} between the No-Risk and At-Risk groups. 
For a category, the group having higher mean rank from MWU Test indicates a stronger association of the category with that group. For the emotion categories that exhibited significant differences (\textit{p} < .05), we further performed bootstrapping to account for Type I error and ensure that the observed differences are not due to chance. Bootstrapping (random sampling) with replacement was performed with 10,000 iterations. Each iteration randomly assigned participants to the At-Risk or No-Risk group. The U-statistic for each iteration was calculated. As a result, we obtain a bootstrap distribution for the U-statistic from which we estimate the significance of the observed statistic. 
\subsection{Genre-Prevalence Analysis}
To further analyse the types of music associated with the tags of emotion categories, we explored genre-related social tags. In order to select the genre-related tags from our data, we collected the results of the multi-stage model proposed by Ferrer et al. \cite{beyondgenre} which assigned tags of Last.fm to different semantic layers namely genre, artist, affect, instrument, etc. In order to understand the underlying genre tag structure and obtain broader genre categories, we employed the approach described by Ferrer et al. \cite{rafael} to cluster genre tags (details in Equation 1 in the Supplementary material). In this, music tags were hierarchically organized revealing taxonomy of music tags by means of latent semantic analysis. The clusters thus obtained were labelled based on the genre-tags constituting the core points of the cluster\cite{treecut}.



For the emotion categories that exhibited significant group differences, the genre tags co-occurring with its tags were used to calculate a user-specific \textit{Genre Prevalence Score} for each genre-tag cluster. The formula used was similar to \textit{Emotion Prevalence Scores} with the change in definition of the following terms: $c$ represents the genre cluster, $T_u$ is the set of  all tracks for user $u$ which have a tag belonging to the particular emotion category and $V_{tg}$ is the set of all genre tags. Finally, we performed a biserial correlation between \textit{Genre Prevalence Scores} for each genre-tag cluster and the users’ risk for depression (represented as a dichotomous variable with 0 = No-Risk; 1 = At-Risk).
\section{Results}\label{Results}
\subsection{Internal Consistency and Criterion Validity}
The Cronbach's alphas for \textit{Unhealthy} scores obtained from HUMS and K10 scores were found to be relatively high at 0.80 and 0.91 respectively. A significant correlation (r=0.55, df=539, p\textless0.001) between \textit{Unhealthy} Score and K10 was found which is in concordance with past research studies in the field \cite{hums}.
Also, in line with previous research \cite{klein,mckenzie}, a significant positive correlation was observed between
K10 score and Neuroticism (r=0.68, p<0.0001) adding to the internal consistency of the data and confirming construct validity.
As can be seen in \figref{boxplot}, the At-Risk group displayed higher mean and median \textit{Unhealthy} score compared to No-Risk while \textit{Healthy} scores were comparable.
\begin{figure}[h]\label{boxplot}
 \centerline{
 \includegraphics[width=0.7\columnwidth]{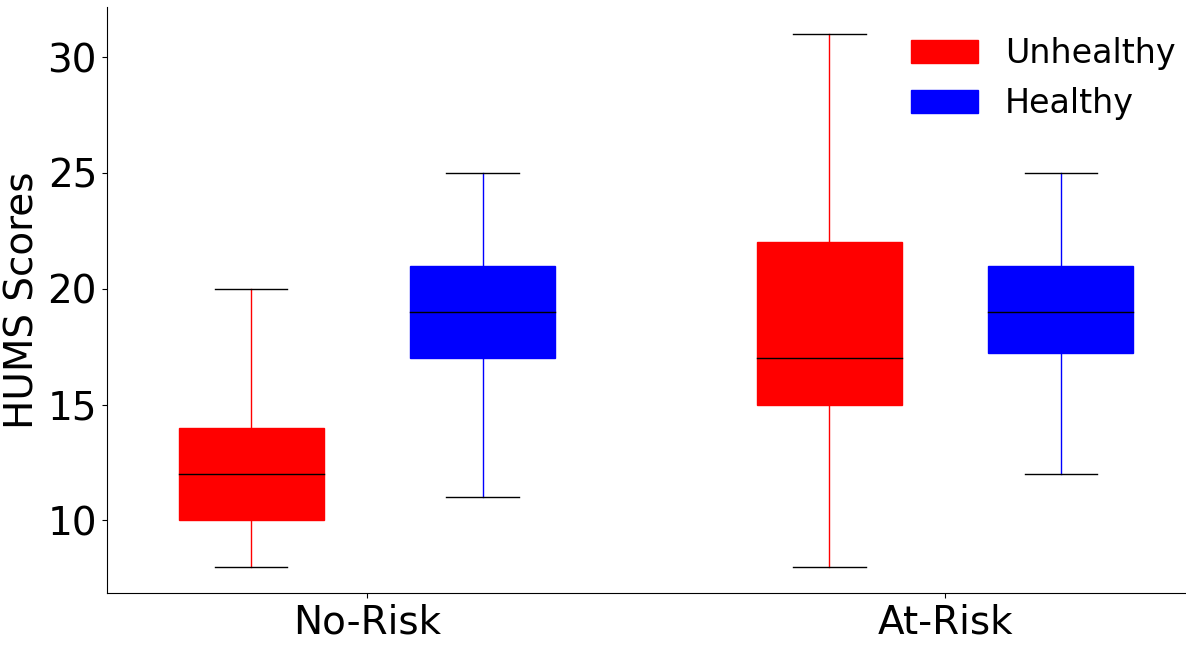}}
 \caption{Boxplot of HUMS scores for No-Risk and At-Risk Groups.}
 \label{boxplot}
\end{figure}
Partial correlations between \textit{Unhealthy}, \textit{Healthy}, and K10 are presented in \tabref{internalconsistency}. K10 scores exhibit significant positive correlation only with \textit{Unhealthy} for both the groups.
The moderate correlation between \textit{Healthy} and \textit{Unhealthy} scores for the No-Risk population indicates that both of these subscales capture a shared element, most likely active music listening.

\begin{table}[h]
\centering
\scalebox{0.9}{
\begin{tabular}{|p{1.3cm} |p{1.0cm}|p{1.3cm}| p{1.0cm}|p{1.3cm}|}
\hline
&\multicolumn{2}{|c|}{No-Risk} & \multicolumn{2}{|c|}{At-Risk} \\
 \hline
 Scales&Healthy& Unhealthy&Healthy&Unhealthy\\
 \hline
 Healthy&1.0&0.36**&1.0&-0.14\\
 Unhealthy&0.36**&1.0&-0.14&1.0\\
 K10& 0.07&0.26**&-0.11&0.22*\\
 \hline
\end{tabular}
}
\caption{Partial Correlation Values between HUMS \& K10. (*p\textless0.01 \& **p\textless0.001)}
\label{internalconsistency}
\end{table}
\subsection{Emotion-based Results}
The data (\textit{t}=$\pm$3,\textit{n}=500) consisted of 3,80,261 social tags. The tag filtering process resulted in a final set of 1254 unique tags (Mean=109, SD = 24 tags per user) which were then mapped onto the VA and VAD emotion spaces. 
Figure 3 in supplementary material displays the tags closest to each of the Emotion Categories based on VA and VAD models.

\begin{figure}[h]\label{gemsboxplot}
 \centerline{
 \includegraphics[height=0.55\columnwidth, width=0.85\columnwidth]{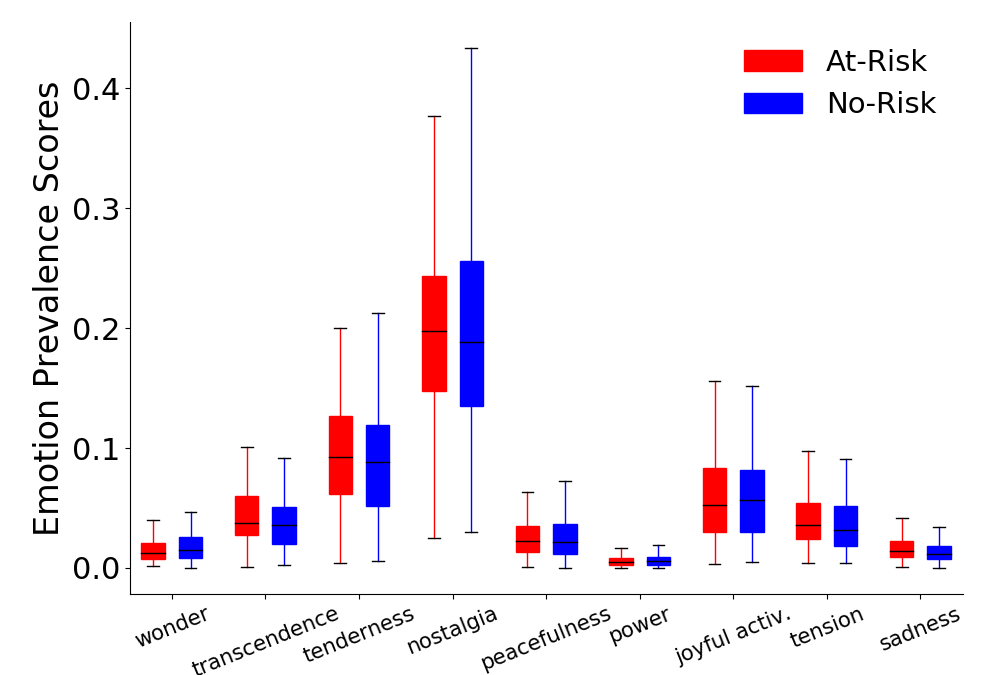}}
 \caption{Boxplot of Emotion Prevalence Scores for No-Risk and At-Risk based on VA.}
 \label{gemsboxplot}
\end{figure}
\figref{gemsboxplot} illustrates the \textit{Emotion Prevalence Scores} for both groups for VA mapping (Supplementary Figure 4 displays the same for VAD, showing a similar distribution). The overall pattern appears similar between both the groups with minor observable differences for the emotion categories \textit{wonder, transcendence, tenderness, tension, and sadness}. \tabref{results} displays the emotion categories that exhibited significant differences between the groups (MWU U-statistic and bootstrap p-values in Table 2 of Supplementary material). The At-Risk group consistently exhibits higher Prevalence Scores in \textit{Sadness} while the No-Risk group vacillates between \textit{Wonder} and \textit{Transcendence}. The most significant difference was observed in \textit{Sadness} (VA model,\textit{t}=$\pm$3,\textit{n}=100) with a significantly greater \textit{Emotion Prevalence Score} for the At-Risk group (Median = 0.0117) than the No-Risk group (Median = 0.0091), U=11414.5, p=0.009. Significant difference was also observed for \textit{Tenderness} with greater \textit{Emotion Prevalence Score} for the At-Risk group (Median = 0.1271) than the No-Risk group (Median = 0.1189), U=11905.0, p=0.04. On the other hand, the \textit{Emotion Prevalence Score} in \textit{Wonder} (VA model,\textit{t}=$\pm$2,\textit{n}=100) was significantly greater for the No-Risk group (Median = 0.0131) than the At-Risk group (Median = 0.0086), U=16270.0, p=0.003. The word-clouds of tags comprising \textit{Sadness} and \textit{Tenderness} are displayed in \figref{sadness} and \figref{tenderness}. A score per tag is computed for each group (Equation 2 in Supplementary material). A rank was assigned to the tag based on the absolute difference of the tag scores between No-Risk \& At-Risk groups. The size of the tag in the word-cloud is directly proportional to its rank in the category. Supplementary figures 5 and 6 depict word-clouds for \textit{Transcendence} and \textit{Wonder}.




\begin{figure}[h!]
  \begin{subfigure}[b]{0.47\linewidth}
    \includegraphics[width=\linewidth]{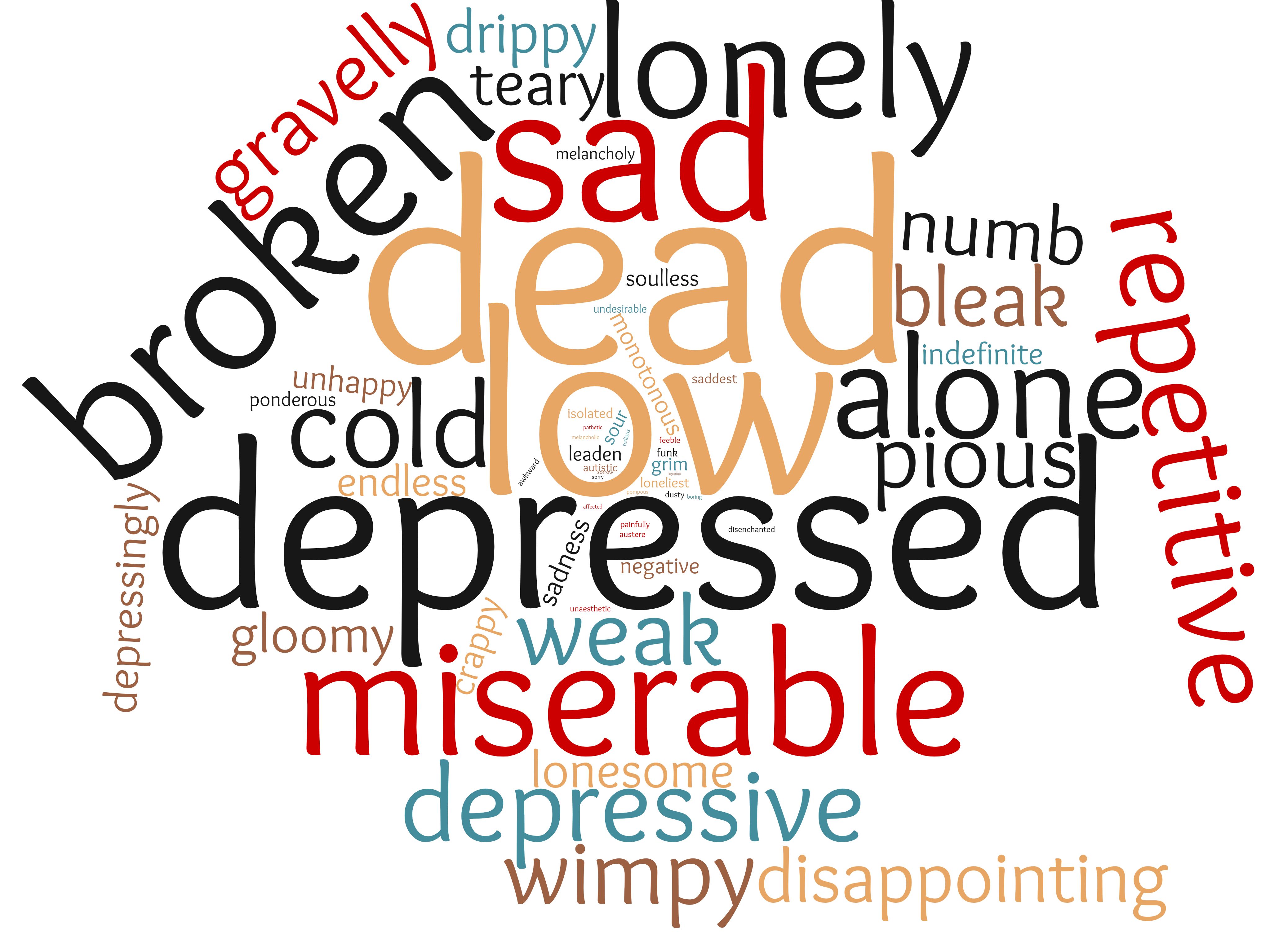}
    \caption{Sadness (VA,$\pm$3,100)}
    \label{sadness}
  \end{subfigure}
  \hfill
  \begin{subfigure}[b]{0.47\linewidth}
    \includegraphics[width=\linewidth]{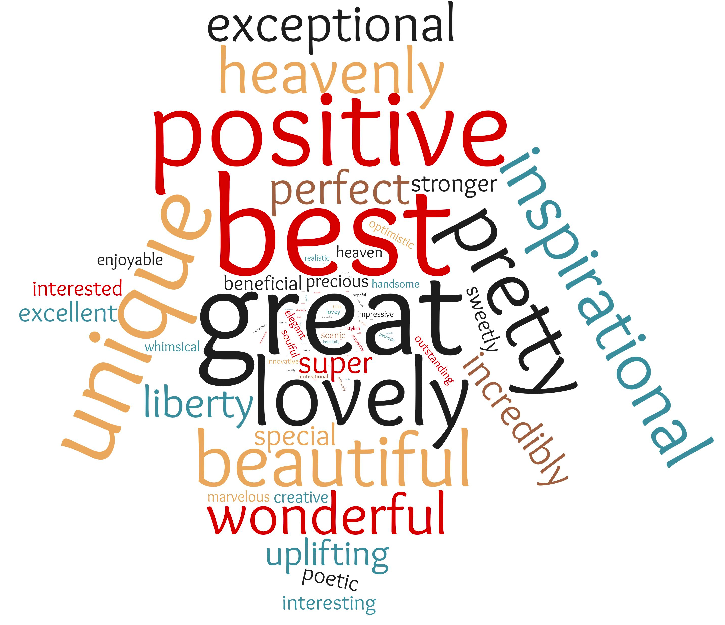}
    \caption{Tenderness (VAD,$\pm$3,500)}
    \label{tenderness}
  \end{subfigure}
  \caption{Wordclouds for emotion categories associated with At-Risk group.}
\end{figure}
We also assessed the predictive power of social tags for risk of depression by classifying participants into At-Risk or No-Risk groups using their tag information (feature details in Equation 4 in Supplementary material). 
The SVM model with 'rbf' kernel (C=2301, gamma=101) gave the best results with a 5-fold cross-validation accuracy of 66.4\%.

\begin{table*}[h]
 \centering
\scalebox{0.9}{
 \begin{tabular}{|p{1.2cm}|p{1.0cm}|p{3.3cm}|p{3.3cm}|p{1.8cm}|p{3.8cm}|}
  \hline
  Group&Top&\multicolumn{2}{|c|}{VAD}&\multicolumn{2}{|c|}{VA}\\\cline{3-6}&Tracks&\textit{t}=$\pm3$&\textit{t}=$\pm2$&\textit{t}=$\pm3$&\textit{t}=$\pm2$\\
  \hline
  &\textit{n}=100&Sadness*&&Sadness**&\\\cline{2-6}
  At-Risk&\textit{n}=200&Sadness*&Sadness*, Tenderness*&Sadness*&Sadness*, Transcendence*\\\cline{2-6}
  &\textit{n}=500&Sadness*, Tenderness*&Tenderness*&Sadness*&Sadness*, Transcendence*\\
  \hline
  &\textit{n}=100&&Transcendence*&&Wonder**\\\cline{2-6}
  No-Risk&\textit{n}=200&Transcendence*&Transcendence*&Wonder*&Wonder**\\\cline{2-6}
  &\textit{n}=500&Transcendence*&Transcendence*&Wonder*&Wonder**\\
  \hline
 \end{tabular}
 }
\footnotesize
 \caption{Emotion Categories with Significant Differences between At-Risk and No-Risk groups.*p\textless0.05; **p\textless0.01}
 \label{results}
\end{table*}
\subsection{Genre-Prevalence Results}
Out of the 5062 tags assigned to the genre layer in \cite{beyondgenre},  94\% (4766) of the tags were present in our data. The clustering of the genre tags resulted in 17 clusters and is displayed in Table 3 of Supplementary material. Figure 7 in Supplementary material displays mean genre prevalence scores between both groups for these 17 clusters. Overall, genre-cluster representing \textit{indie-,alternative-pop/rock} represented by Cluster 4 is predominant in both groups. Genre prevalence scores were then evaluated specific to the tracks associated with the emotion categories that exhibited most significant group differences, that is, \textit{Wonder} and \textit{Sadness} (VA model, \textit{t}=$\pm$3, \textit{n}=100). For \textit{Sadness}-specific tracks, the highest correlation (r=0.2, p<0.01) was observed between the Genre-Prevalence scores in the cluster representing \textit{neo psychedelic-, avant garde-, dream-pop} and K-10 scores. Also, genre clusters representing \textit{electronic rock} (r=0.17, p<0.01), \textit{indie-, alternative-pop/rock} (r=0.12, p<0.05), and \textit{world music} (r=-0.11, p<0.05) demonstrated significant correlations for \textit{Tenderness}. For \textit{Wonder} (VA model, \textit{t}=$\pm$2, \textit{n}=100), the K-10 scores exhibited significant negative correlation with \textit{Genre Prevalence Scores} of clusters representing \textit{black metal} (r=-0.11, p<0.05) and \textit{neo-progressive rock} (r=-0.13, p<0.05).

\section{Discussion}
This study is the first of its kind to examine the association between risk for depression and social tags related to music listening habits as they occur naturally as opposed to self-reported or lab-based studies.
A clear difference in the music listening profiles was observed between the At-Risk group and the No-Risk group, particularly in terms of the emotional content of the tags. \textit{Sadness} was significantly more prevalent in the At-Risk group and the word-cloud of sadness was highly illustrative of other low-arousal, low-valence emotions such as \textit{dead}, \textit{low}, \textit{depressed, miserable, broken}, and \textit{lonely}. The stronger association of the At-Risk group with sadness is in concordance with the past research studies in the field \cite{garrido} and confirms our hypothesis. The At-Risk group is attracted to music that reflects and resonates with their internal state. Whether this provides emotional consolation as an adaptive resource or whether it only worsens repetitive negative feelings and fuels rumination, remains an open question. Nonetheless, statistically, such listening style can be seen as a highly predictive factor of psychological distress.

In addition, \textit{Tenderness}, which represents low-arousal and high-valence, was also more prevalent in the At-Risk group, especially for shorter-term ($\pm$ 2 months) music listening habits. \textit{Tenderness} appears to be more significant in the shorter time period in addition to \textit{Sadness}, possibly indicating that At-Risk people tend to oscillate between positive and negative states within a general state of low arousal. These findings appear to be very much in line with the results found by Houben et al. \cite{houben2015relation} who found high levels of emotional inertia and emotional variability to be linked with depression and ill-being. The consistent results related to \textit{Sadness} in our study reflect the overall inert states in which the At-Risk tend to be. On the other hand, the \textit{Tenderness} results reflect their tendency to jump to positive affective states while retaining low arousal, thereby demonstrating emotional variability. Furthermore, the omission of the \textit{Dominance} dimension causes most of the tags to shift from \textit{Tenderness} to \textit{Transcendence} and \textit{Transcendence} to \textit{Wonder}, which explains the results in a reversal of the group association as evidenced in the results. Nevertheless, \textit{Sadness} appears to be the predominant state as it is largely consistent for $\pm$3 months as well as for $\pm$2 months of music listening histories. 

The At-Risk group also exhibited a tendency to gravitate towards music with genre tags such as \textit{neo-psychedelic-}, \textit{avant garde-}, \textit{dream-pop} co-occurring with \textit{Sadness}. Such genres are characterized by ethereal-sounding mixtures that often result in a wall of sound comprising electronic textures with obscured vocals. 
Similarly, the genres co-occurring with \textit{Tenderness} (VAD model) or \textit{Transcendence} (VA model) comprise similar mixtures with heavy synthesizer-based sounds (such as mellotron) which result in sounds that seem otherwordly. Such out-of-this world soundscapes have been also associated with transcendent druggy and mystical imagery and immersive experiences\cite{goddard2013resonances}. These results strengthen the claim that depression may foster musical immersion as an escape from a reality that is perceived to be adverse. This is somewhat in line with prior research that has linked depression with the use of music for avoidant coping\cite{miranda2009music}.
On the other hand, music listening history of the No-Risk group was characterized by an inclination to listen to music tagged by positive valence and higher arousal as characterized by \textit{Wonder} with a predilection for \textit{Heavy metal} and \textit{Progressive Rock} genres. 

The use of only single word tags in the third stage of tag filtering is one limitation of this study which is due to lack of compatibility of the word emotion induction model with multi-word tags. Our results could potentially be extended to find significant differences in emotional concepts after considering multi-word social tags. We achieve a decent classification accuracy of 66.4\% which is significantly above the chance level which indicate that social tags indeed may be indicative of At-Risk behavior. This may further be improved by considering additional descriptors of music such as acoustic features and lyrical content of the tracks. Another future direction is to incorporate the temporal evolution of these emotion categories in the listening histories to characterize depression, since past research suggests depression to be a result of gradual development of daily emotional experiences \cite{psych}. This study is intended to be one of many to come that will be helpful in early detection of depression and other potential mental disorders in individuals using their digital music footprints. 
\bibliography{ISMIRtemplate}

\begin{thebibliography}{10}
\providecommand{\url}[1]{#1}
\csname url@samestyle\endcsname
\providecommand{\newblock}{\relax}
\providecommand{\bibinfo}[2]{#2}
\providecommand{\BIBentrySTDinterwordspacing}{\spaceskip=0pt\relax}
\providecommand{\BIBentryALTinterwordstretchfactor}{4}
\providecommand{\BIBentryALTinterwordspacing}{\spaceskip=\fontdimen2\font plus
\BIBentryALTinterwordstretchfactor\fontdimen3\font minus
  \fontdimen4\font\relax}
\providecommand{\BIBforeignlanguage}[2]{{%
\expandafter\ifx\csname l@#1\endcsname\relax
\typeout{** WARNING: IEEEtran.bst: No hyphenation pattern has been}%
\typeout{** loaded for the language `#1'. Using the pattern for}%
\typeout{** the default language instead.}%
\else
\language=\csname l@#1\endcsname
\fi
#2}}
\providecommand{\BIBdecl}{\relax}
\BIBdecl

\bibitem{world2017depression}
W.~H. Organization \emph{et~al.}, ``Depression and other common mental
  disorders: global health estimates,'' World Health Organization, Tech. Rep.,
  2017.

\bibitem{munmun}
M.~De~Choudhury, M.~Gamon, S.~Counts, and E.~Horvitz, ``Predicting depression
  via social media,'' in \emph{Seventh international AAAI conference on weblogs
  and social media}, 2013.

\bibitem{copper}
G.~Coppersmith, M.~Dredze, and C.~Harman, ``Quantifying mental health signals
  in twitter,'' in \emph{Proceedings of the workshop on computational
  linguistics and clinical psychology: From linguistic signal to clinical
  reality}, 2014, pp. 51--60.

\bibitem{munmun_14(2)}
M.~De~Choudhury, S.~Counts, E.~J. Horvitz, and A.~Hoff, ``Characterizing and
  predicting postpartum depression from shared facebook data,'' in
  \emph{Proceedings of the 17th ACM conference on Computer supported
  cooperative work \& social computing}, 2014, pp. 626--638.

\bibitem{munmun_16}
M.~De~Choudhury, E.~Kiciman, M.~Dredze, G.~Coppersmith, and M.~Kumar,
  ``Discovering shifts to suicidal ideation from mental health content in
  social media,'' in \emph{Proceedings of the 2016 CHI conference on human
  factors in computing systems}, 2016, pp. 2098--2110.

\bibitem{reece}
A.~G. Reece and C.~M. Danforth, ``Instagram photos reveal predictive markers of
  depression,'' \emph{EPJ Data Science}, vol.~6, no.~1, pp. 1--12, 2017.

\bibitem{baltazar}
M.~Baltazar and S.~Saarikallio, ``Toward a better understanding and
  conceptualization of affect self-regulation through music: A critical,
  integrative literature review,'' \emph{Psychology of Music}, vol.~44, no.~6,
  pp. 1500--1521, 2016.

\bibitem{nave}
G.~Nave, J.~Minxha, D.~M. Greenberg, M.~Kosinski, D.~Stillwell, and
  J.~Rentfrow, ``Musical preferences predict personality: evidence from active
  listening and facebook likes,'' \emph{Psychological Science}, vol.~29, no.~7,
  pp. 1145--1158, 2018.

\bibitem{qiu}
L.~Qiu, J.~Chen, J.~Ramsay, and J.~Lu, ``Personality predicts words in favorite
  songs,'' \emph{Journal of Research in Personality}, vol.~78, pp. 25--35,
  2019.

\bibitem{litlink}
K.~S. McFerran, S.~Garrido, and S.~Saarikallio, ``A critical interpretive
  synthesis of the literature linking music and adolescent mental health,''
  \emph{Youth \& Society}, vol.~48, no.~4, pp. 521--538, 2016.

\bibitem{rumin}
S.~Garrido, T.~Eerola, and K.~McFerran, ``Group rumination: Social interactions
  around music in people with depression,'' \emph{Frontiers in psychology},
  vol.~8, p. 490, 2017.

\bibitem{mcferran}
K.~S. McFerran, ``Contextualising the relationship between music, emotions and
  the well-being of young people: A critical interpretive synthesis,''
  \emph{Musicae Scientiae}, vol.~20, no.~1, pp. 103--121, 2016.

\bibitem{hums}
S.~Saarikallio, C.~Gold, and K.~McFerran, ``Development and validation of the
  healthy-unhealthy music scale,'' \emph{Child and adolescent mental health},
  vol.~20, no.~4, pp. 210--217, 2015.

\bibitem{ragarwal}
R.~Agarwal, R.~Singh, S.~Saarikallio, K.~McFerran, and V.~Alluri, ``Mining
  mental states using music associations,'' \emph{depression}, vol.~2, p.~6,
  2019.

\bibitem{stewart}
J.~Stewart, S.~Garrido, C.~Hense, and K.~McFerran, ``Music use for mood
  regulation: self-awareness and conscious listening choices in young people
  with tendencies to depression,'' \emph{Frontiers in psychology}, vol.~10, p.
  1199, 2019.

\bibitem{greenberg2017social}
D.~M. Greenberg and P.~J. Rentfrow, ``Music and big data: a new frontier,''
  \emph{Current opinion in behavioral sciences}, vol.~18, pp. 50--56, 2017.

\bibitem{saari}
P.~Saari and T.~Eerola, ``Semantic computing of moods based on tags in social
  media of music,'' \emph{IEEE Transactions on Knowledge and Data Engineering},
  vol.~26, no.~10, pp. 2548--2560, 2013.

\bibitem{laurier}
C.~Laurier, M.~Sordo, J.~Serra, and P.~Herrera, ``Music mood representations
  from social tags.'' in \emph{International Society for Music Information
  Retrieval (ISMIR) Conference}, 2009, pp. 381--386.

\bibitem{gupta}
K.~Gupta, N.~Sachdeva, and V.~Pudi, ``Explicit modelling of the implicit short
  term user preferences for music recommendation,'' in \emph{European
  Conference on Information Retrieval}.\hskip 1em plus 0.5em minus 0.4em\relax
  Springer, 2018, pp. 333--344.

\bibitem{polignano}
M.~Polignano, P.~Basile, M.~de~Gemmis, and G.~Semeraro, ``Social tags and
  emotions as main features for the next song to play in automatic playlist
  continuation,'' in \emph{Adjunct Publication of the 27th Conference on User
  Modeling, Adaptation and Personalization}, 2019, pp. 235--239.

\bibitem{k10}
R.~C. Kessler, G.~Andrews, L.~J. Colpe, E.~Hiripi, D.~K. Mroczek, S.-L.
  Normand, E.~E. Walters, and A.~M. Zaslavsky, ``Short screening scales to
  monitor population prevalences and trends in non-specific psychological
  distress,'' \emph{Psychological medicine}, vol.~32, no.~6, pp. 959--976,
  2002.

\bibitem{sakka}
L.~S. Sakka and P.~N. Juslin, ``Emotion regulation with music in depressed and
  non-depressed individuals: Goals, strategies, and mechanisms,'' \emph{Music
  \& Science}, vol.~1, p. 2059204318755023, 2018.

\bibitem{topo}
M.~B. Donnellan, F.~L. Oswald, B.~M. Baird, and R.~E. Lucas, ``The mini-ipip
  scales: tiny-yet-effective measures of the big five factors of personality.''
  \emph{Psychological assessment}, vol.~18, no.~2, p. 192, 2006.

\bibitem{valence_shifters}
L.~Polanyi and A.~Zaenen, ``Contextual valence shifters,'' in \emph{Computing
  attitude and affect in text: Theory and applications}.\hskip 1em plus 0.5em
  minus 0.4em\relax Springer, 2006, pp. 1--10.

\bibitem{dimension1}
R.~Trnka, A.~La{\v{c}}ev, K.~Balcar, M.~Ku{\v{s}}ka, and P.~Tavel, ``Modeling
  semantic emotion space using a 3d hypercube-projection: an innovative
  analytical approach for the psychology of emotions,'' \emph{Frontiers in
  psychology}, vol.~7, p. 522, 2016.

\bibitem{eurola}
T.~Eerola and J.~K. Vuoskoski, ``A comparison of the discrete and dimensional
  models of emotion in music,'' \emph{Psychology of Music}, vol.~39, no.~1, pp.
  18--49, 2011.

\bibitem{dimension2}
Z.~Zhu, J.~Li, X.~Deng, Y.~Hu \emph{et~al.}, ``An improved three-dimensional
  model for emotion based on fuzzy theory,'' \emph{Journal of Computer and
  Communications}, vol.~6, no.~08, p. 101, 2018.

\bibitem{VAmodel}
J.~A. Russell, ``A circumplex model of affect.'' \emph{Journal of personality
  and social psychology}, vol.~39, no.~6, p. 1161, 1980.

\bibitem{wei}
S.~Buechel and U.~Hahn, ``Word emotion induction for multiple languages as a
  deep multi-task learning problem,'' in \emph{Proceedings of the 2018
  Conference of the North American Chapter of the Association for Computational
  Linguistics: Human Language Technologies, Volume 1 (Long Papers)}, 2018, pp.
  1907--1918.

\bibitem{vad}
M.~M. Bradley and P.~J. Lang, ``Measuring emotion: the self-assessment manikin
  and the semantic differential,'' \emph{Journal of behavior therapy and
  experimental psychiatry}, vol.~25, no.~1, pp. 49--59, 1994.

\bibitem{paasi_2}
P.~Saari, M.~Barthet, G.~Fazekas, T.~Eerola, and M.~Sandler, ``Semantic models
  of musical mood: Comparison between crowd-sourced and curated editorial
  tags,'' in \emph{2013 IEEE International Conference on Multimedia and Expo
  Workshops (ICMEW)}.\hskip 1em plus 0.5em minus 0.4em\relax IEEE, 2013, pp.
  1--6.

\bibitem{musicvad}
M.~Buccoli, M.~Zanoni, G.~Fazekas, A.~Sarti, and M.~B. Sandler, ``A
  higher-dimensional expansion of affective norms for english terms for music
  tagging.'' in \emph{ISMIR}, 2016, pp. 316--322.

\bibitem{musicvad2}
F.~H. Rachman, R.~Sarno, and C.~Fatichah, ``Music emotion classification based
  on lyrics-audio using corpus based emotion.'' \emph{International Journal of
  Electrical \& Computer Engineering (2088-8708)}, vol.~8, no.~3, 2018.

\bibitem{fasttext}
P.~Bojanowski, E.~Grave, A.~Joulin, and T.~Mikolov, ``Enriching word vectors
  with subword information,'' \emph{Transactions of the Association for
  Computational Linguistics}, vol.~5, pp. 135--146, 2017.

\bibitem{Warriner2013}
A.~B. Warriner, V.~Kuperman, and M.~Brysbaert, ``Norms of valence, arousal, and
  dominance for 13,915 english lemmas,'' \emph{Behavior research methods},
  vol.~45, no.~4, pp. 1191--1207, 2013.

\bibitem{gems}
M.~Zentner, D.~Grandjean, and K.~R. Scherer, ``Emotions evoked by the sound of
  music: characterization, classification, and measurement.'' \emph{Emotion},
  vol.~8, no.~4, p. 494, 2008.

\bibitem{gemsvsVAD}
J.~K. Vuoskoski and T.~Eerola, ``Domain-specific or not? the applicability of
  different emotion models in the assessment of music-induced emotions,'' in
  \emph{Proceedings of the 10th international conference on music perception
  and cognition}, 2010, pp. 196--199.

\bibitem{beyondgenre}
R.~Ferrer and T.~Eerola, ``Looking beyond genres: Identifying meaningful
  semantic layers from tags in online music collections,'' in \emph{2011 10th
  International Conference on Machine Learning and Applications and Workshops},
  vol.~2.\hskip 1em plus 0.5em minus 0.4em\relax IEEE, 2011, pp. 112--117.

\bibitem{rafael}
------, ``Semantic structures of timbre emerging from social and acoustic
  descriptions of music,'' \emph{EURASIP Journal on Audio, Speech, and Music
  Processing}, vol. 2011, no.~1, p.~11, 2011.

\bibitem{treecut}
P.~Langfelder, B.~Zhang, and S.~Horvath, ``Defining clusters from a
  hierarchical cluster tree: the dynamic tree cut package for r,''
  \emph{Bioinformatics}, vol.~24, no.~5, pp. 719--720, 2008.

\bibitem{klein}
D.~N. Klein, R.~Kotov, and S.~J. Bufferd, ``Personality and depression:
  explanatory models and review of the evidence,'' \emph{Annual review of
  clinical psychology}, vol.~7, p. 269, 2011.

\bibitem{mckenzie}
S.~H. McKenzie, U.~W. Jayasinghe, M.~Fanaian, M.~Passey, D.~Lyle, G.~P. Davies,
  and M.~F. Harris, ``Socio-demographic factors, behaviour and personality:
  associations with psychological distress,'' \emph{European journal of
  preventive cardiology}, vol.~19, no.~2, pp. 250--257, 2012.

\bibitem{garrido}
S.~Garrido and E.~Schubert, ``Music and people with tendencies to depression,''
  \emph{Music Perception: An Interdisciplinary Journal}, vol.~32, no.~4, pp.
  313--321, 2015.

\bibitem{houben2015relation}
M.~Houben, W.~Van Den~Noortgate, and P.~Kuppens, ``The relation between
  short-term emotion dynamics and psychological well-being: A meta-analysis.''
  \emph{Psychological bulletin}, vol. 141, no.~4, p. 901, 2015.

\bibitem{goddard2013resonances}
M.~Goddard, B.~Halligan, and N.~Spelman, \emph{Resonances: noise and
  contemporary music}.\hskip 1em plus 0.5em minus 0.4em\relax A\&C Black, 2013.

\bibitem{miranda2009music}
D.~Miranda and M.~Claes, ``Music listening, coping, peer affiliation and
  depression in adolescence,'' \emph{Psychology of music}, vol.~37, no.~2, pp.
  215--233, 2009.

\bibitem{psych}
D.~Miranda, P.~Gaudreau, R.~Debrosse, J.~Morizot, and L.~J. Kirmayer, ``Music
  listening and mental health: Variations on internalizing psychopathology,''
  \emph{Music, health, and wellbeing}, pp. 513--529, 2012.

\end{thebibliography}


\begin{thebibliography}{1}
\providecommand{\url}[1]{#1}
\csname url@samestyle\endcsname
\providecommand{\newblock}{\relax}
\providecommand{\bibinfo}[2]{#2}
\providecommand{\BIBentrySTDinterwordspacing}{\spaceskip=0pt\relax}
\providecommand{\BIBentryALTinterwordstretchfactor}{4}
\providecommand{\BIBentryALTinterwordspacing}{\spaceskip=\fontdimen2\font plus
\BIBentryALTinterwordstretchfactor\fontdimen3\font minus
  \fontdimen4\font\relax}
\providecommand{\BIBforeignlanguage}[2]{{%
\expandafter\ifx\csname l@#1\endcsname\relax
\typeout{** WARNING: IEEEtran.bst: No hyphenation pattern has been}%
\typeout{** loaded for the language `#1'. Using the pattern for}%
\typeout{** the default language instead.}%
\else
\language=\csname l@#1\endcsname
\fi
#2}}
\providecommand{\BIBdecl}{\relax}
\BIBdecl

\bibitem{gems}
M.~Zentner, D.~Grandjean, and K.~R. Scherer, ``Emotions evoked by the sound of
  music: characterization, classification, and measurement.'' \emph{Emotion},
  vol.~8, no.~4, p. 494, 2008.

\bibitem{rafael}
R.~Ferrer and T.~Eerola, ``Semantic structures of timbre emerging from social
  and acoustic descriptions of music,'' \emph{EURASIP Journal on Audio, Speech,
  and Music Processing}, vol. 2011, no.~1, p.~11, 2011.

\bibitem{treecut}
P.~Langfelder, B.~Zhang, and S.~Horvath, ``Defining clusters from a
  hierarchical cluster tree: the dynamic tree cut package for r,''
  \emph{Bioinformatics}, vol.~24, no.~5, pp. 719--720, 2008.

\end{thebibliography}

%
%
%
%

\end{document}


%
%

\section*{Supplementary Material of "Tag2Risk: Harnessing Social Music Tags for Characterizing Depression Risk"}



\vspace*{5px}


\begin{figure}[H]
 \centerline{
 \includegraphics[width=0.95\columnwidth]{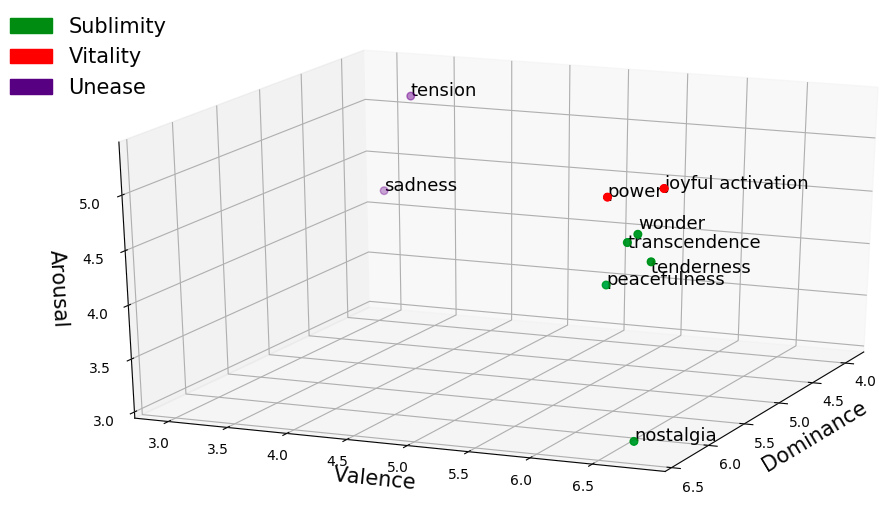}}
 \caption{Scatter plot of the Emotion Categories projected onto the Valence-Arousal-Dominance (VAD) space. The points have been color coded according to the overarching 3 categories of GEMS\cite{gems} as displayed in \tabref{gemsloading}. The VAD values for the 40 emotion terms as presented in \tabref{gemsloading} were obtained from the Tag Emotion Induction model. These were then weighted and summed according to their respective factor loadings to finally obtain VAD values for each of the 9 emotion categories.}
 \label{gemsvad}
\end{figure}

\begin{figure}[H]
 \centerline{
 \includegraphics[width=0.95\columnwidth]{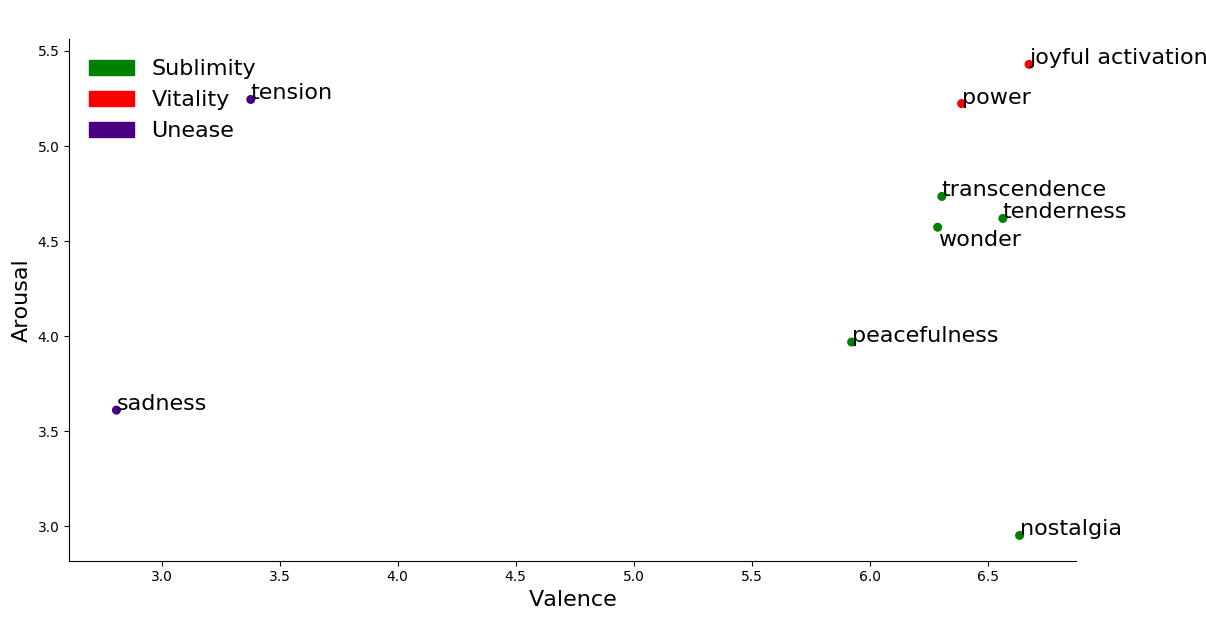}}
 \caption{Scatter plot of the Emotion Categories projected onto the Valence-Arousal (VA) space. The points have been color coded according to the overarching 3 categories of GEMS\cite{gems} as displayed in \tabref{gemsloading}. The VA values for the 40 emotion terms as presented in \tabref{gemsloading} were obtained from the Tag Emotion Induction model. These were then weighted and summed according to their respective factor loadings to finally obtain VA values for each of the 9 emotion categories.}
 \label{gemsva}
\end{figure}


\begin{figure}[H]
 \centerline{
 \includegraphics[width=\textwidth]{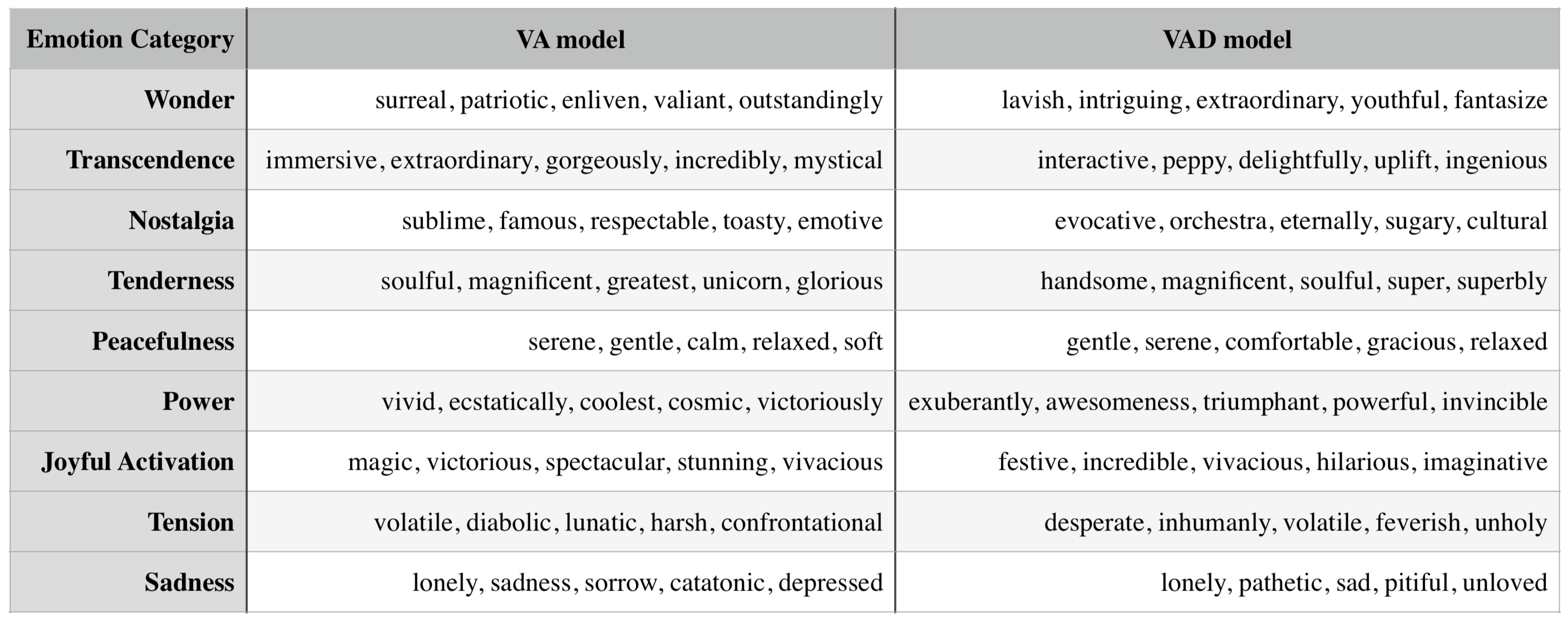}}
 \caption{Tags closest to each of the Emotion Categories in VA/VAD space. Each of the tags are assigned the emotion category based on the proximity in the VA(or VAD) space as evaluated by the euclidean distance from the projected position of the emotion category term.}
 \label{gemstags}
\end{figure}

\begin{figure}[H]
 \centerline{
 \includegraphics[width=\columnwidth]{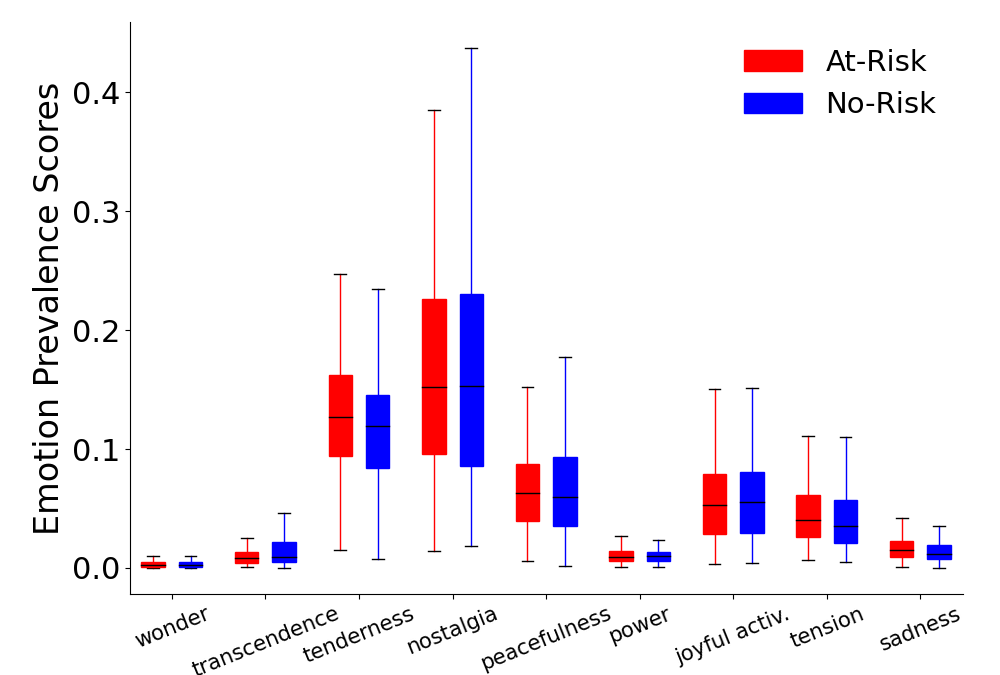}}
 \caption{Boxplot of \textit{Emotion Prevalence Scores} for No-risk and At-risk based on VAD space. These \textit{Emotion Prevalence Scores} were computed per user and represent the presence of tags belonging to that particular emotion category in the user's listening history.}
 \label{bpVAD}
\end{figure}




\begin{figure}[H]
  \begin{subfigure}[b]{0.46\linewidth}
    \includegraphics[width=\linewidth]{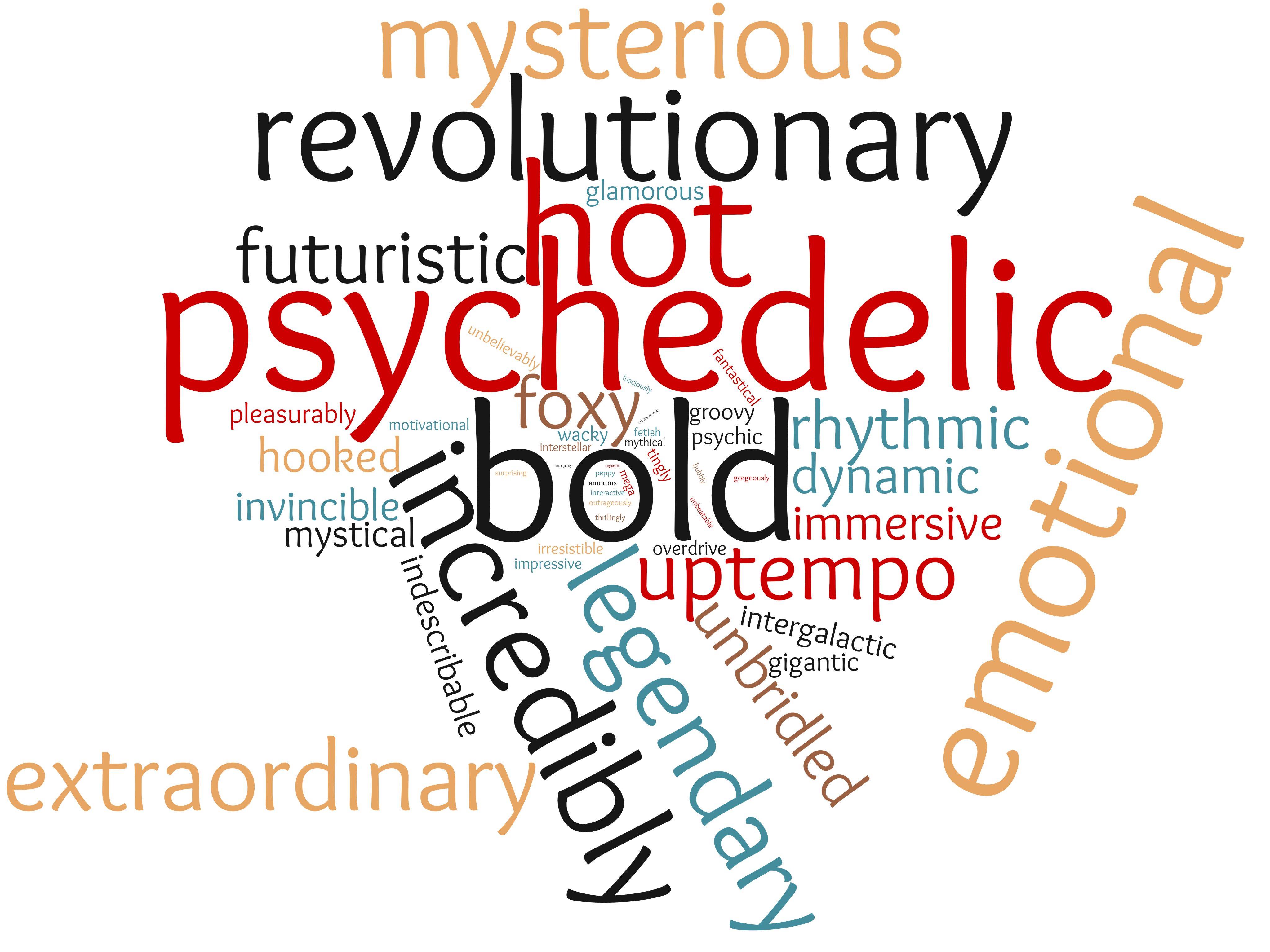}
    \caption{Wordcloud for \textit{transcendence} emotion category based on VA space (t=$\pm$2,n=200) which had higher association with the At-Risk group.}
    \label{transva}
  \end{subfigure}
  \hfill
  \begin{subfigure}[b]{0.46\linewidth}
    \includegraphics[width=\linewidth]{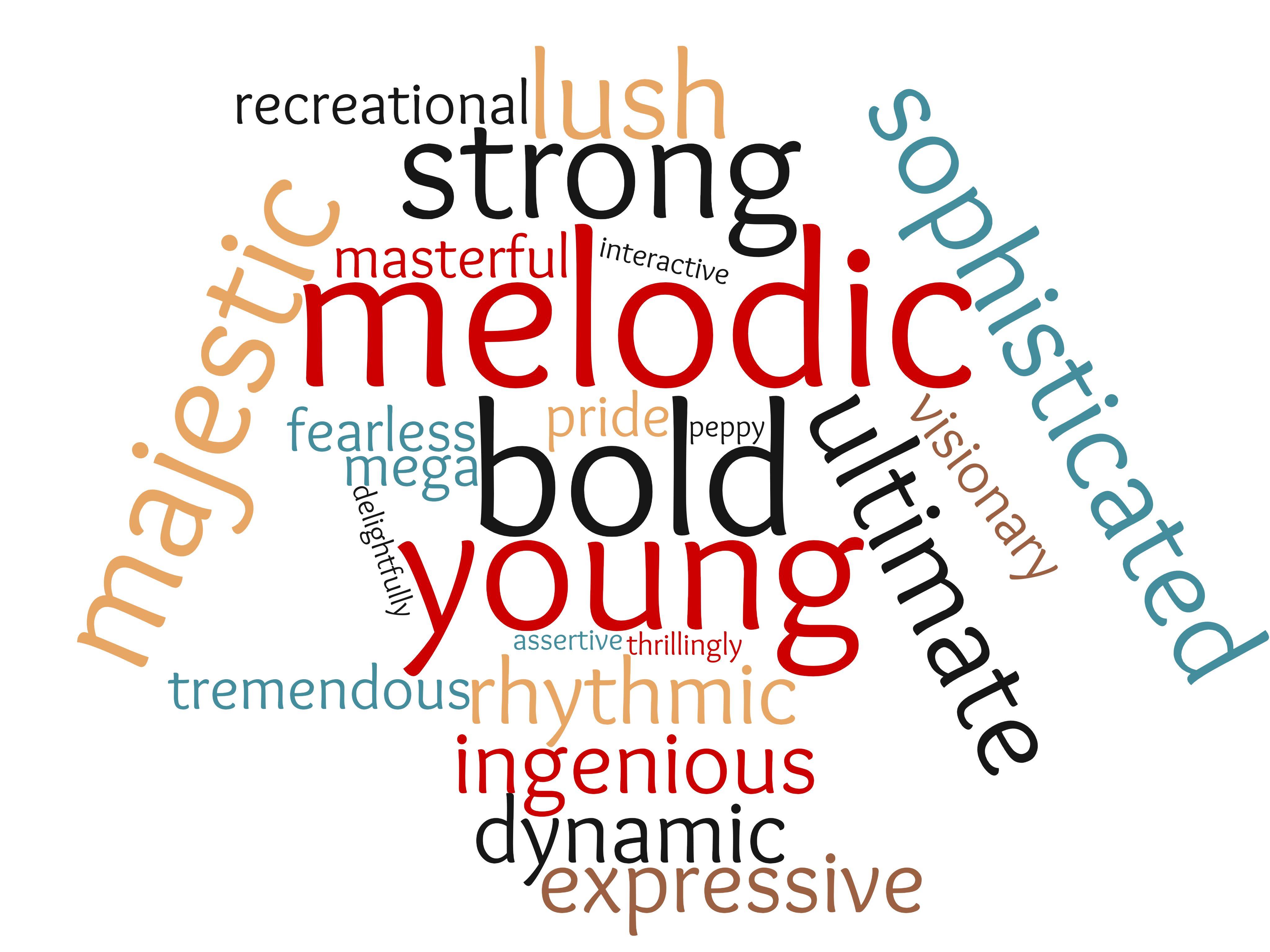}
    \caption{Wordcloud for \textit{transcendence} emotion category based on VAD space (t=$\pm$2,n=200) which had higher association with the No-Risk group.}
    \label{transvad}
  \end{subfigure}
  \caption{Wordclouds for \textit{transcendence} emotion category. The size of the tag in the word-cloud is directly proportional to its rank in the category as computed in \eqnref{tagscore}.}
\end{figure}

\begin{figure}[H]
 \centerline{
 \includegraphics[width=0.46\textwidth]{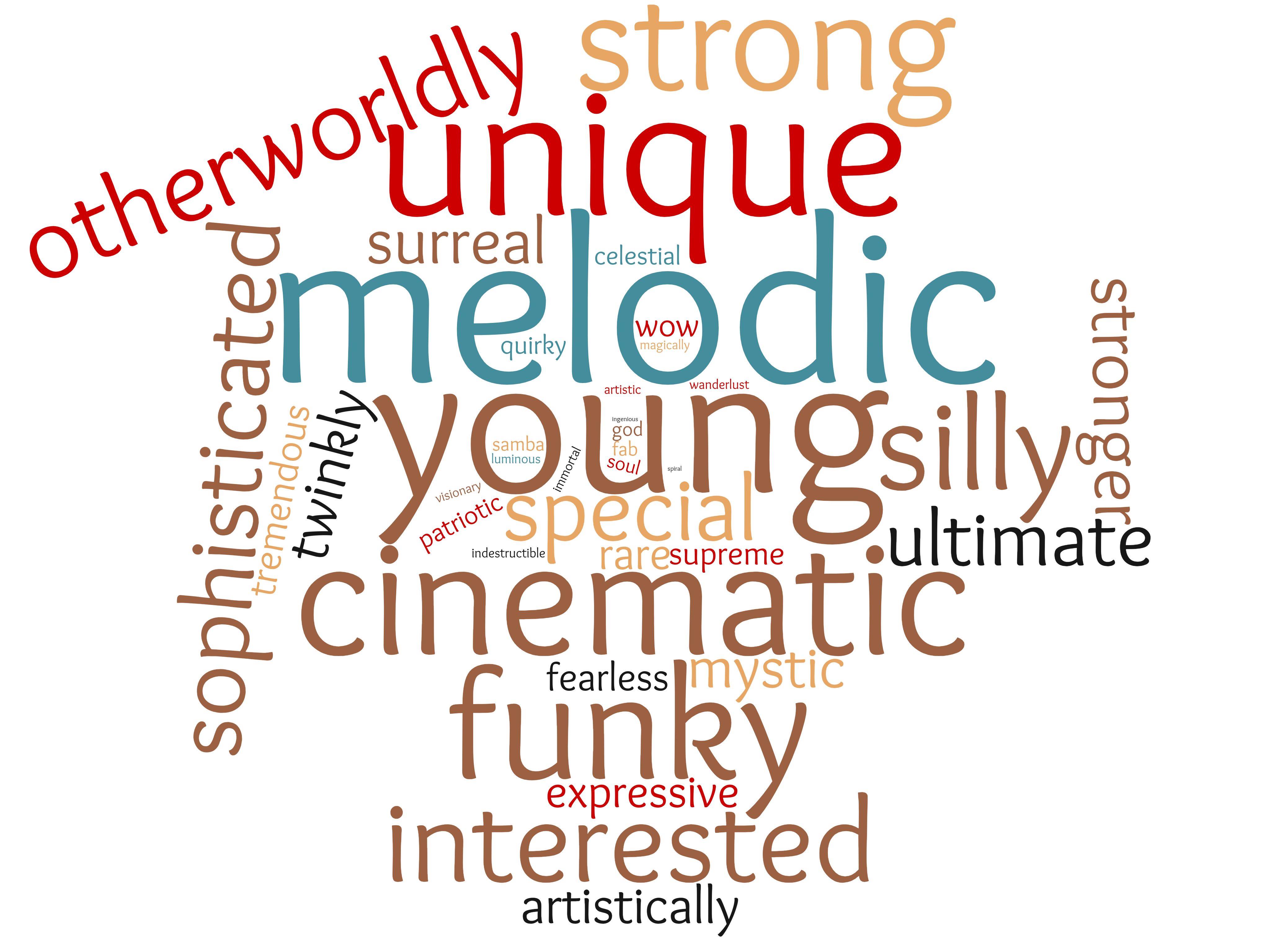}}
 \caption{Wordcloud for \textit{wonder} emotion category based on VA space (t=$\pm$2,n=100) which had higher association with the No-Risk group. The size of the tag in the word-cloud is directly proportional to its rank in the category as computed in \eqnref{tagscore}.}
 \label{wonderva}
\end{figure}



\begin{figure}[H]
 \centerline{
 \includegraphics[width=0.7\columnwidth]{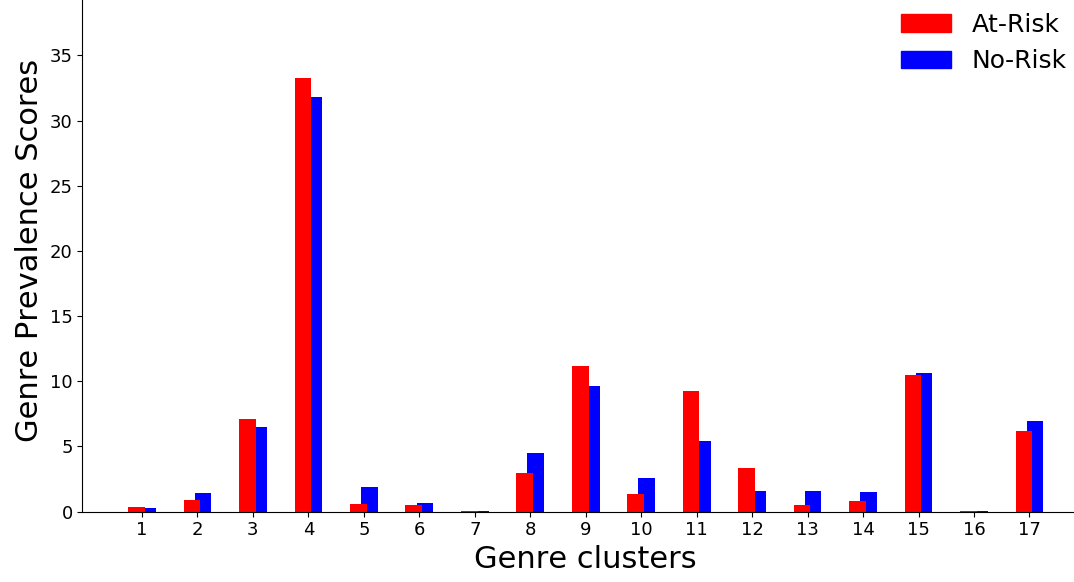}}
 \caption{Mean \textit{Genre Prevalence Scores} for At-Risk and No-Risk Groups for each of the 17 genre clusters. The most representative tags along with the corresponding cluster labels can be found in \tabref{clusters}.}
 \label{genredist}
\end{figure}

\subsection*{Genre-tag clustering}
In order to understand the underlying genre tag structure and obtain broader genre categories, we employed the approach described by Ferrer et al. \cite{rafael} to cluster genre tags. The genre-tags were structured using a vector-based semantic representation in three stages. First, we constructed a \textit{Term-Document Matrix}  $X = x_{ij}$. Every track 'i' corresponds to a “Document” and genre-tag 'j' to a “Term”. This matrix captures tag co-occurrences for each track. Second, similarity coefficients for each pair of genre-tags were computed based on their presence($x_{ij}=1$) or absence($x_{ij}=0$) as shown in \eqnref{genrecluster}.
\begin{equation}\label{genrecluster}
    D_{ij} = \dfrac{ad}{\sqrt{(a+b)(a+c)(b+d)(c+d)}}
\end{equation}
where \textit{a} equals the number of (1,1) co-occurrences, \textit{b} = (1,0), \textit{c} = (0,1) and \textit{d} = (0,0) respectively. Finally, hierarchical clustering was performed on the similarity matrix using Ward's minimum variance method. The dynamic tree cut algorithm was used to find the optimal cut height as it offers advantages over the constant height cutoff values which may exhibit sub-optimal performance on complicated dendrograms \cite{treecut}. The dendrogram was then cut at the obtained height to extract genre-tag clusters. The clusters thus obtained were labelled based on the genre-tags constituting the core points of the cluster.

\subsection*{Tag ranking in wordcloud}
A score for each tag per group is computed as displayed in Equation \eqref{tagscore} by summing the product of the normalised tag-weight and the normalised track playcount for each of tracks in the listening history of all the users in that group.
\begin{equation}\label{tagscore}
G_{g,t}=\frac{\sum_{u\epsilon{U\textsubscript{g}}}\sum_{j\epsilon{V\textsubscript{tr}}}\left(N_{j,t}\times tr\textsubscript{u,j}\right)}{\sum_{u\epsilon{U\textsubscript{g}}}\sum_{i\epsilon{T\textsubscript{u}}} tr \textsubscript{u,i}}
\end{equation}
where
\begin{equation}\label{n_tagweight}
N_{j,t}=\frac{tw\textsubscript{j,t}}{\sum_{l\epsilon{V\textsubscript{tg}}}tw\textsubscript{j,l}}
\end{equation}
$t$ : tag\\
$g$ : group (No-Risk or At-risk)\\
$U_g$ : all users in group $g$\\
$N_{j,t}$ : the association of track $j$ with $t$ \\ $T_u$ : all tracks for user $u$\\ $V_{tg}$ : all tags obtained after tag filtering\\ $V_{tr}$ : all tracks having at least one tag from $V_{tg}$\\ $tr_{u,i}$ : playcount of track $i$ for user $u$\\ $tw_{j,l}$ : tag weight of tag $l$ for track $j$ \\

The rank is assigned to the tag based on the absolute difference of the tag scores between No-Risk \& At-Risk groups. The size of the tag in the word-cloud is directly proportional to its rank in the category.

\subsection*{Classification using emotion related tags}
We assess the predictive power of social tags for risk for depression by classifying participants into At-Risk or No-risk groups using their tag information. The tags used are the emotion related tags. To represent tag information for each user, a 300 dimensional vector is computed representing the weighted average of the fasttext embeddings of the tags in the participant's listening history as in \eqref{embedding}.

\begin{equation}\label{embedding}
E_u = \sum_{t \in V_{tg}} (ts_{u,t} * ft_t )
\end{equation}
where,
\begin{equation}
ts_{u,t}=\frac{\sum_{j\epsilon{V\textsubscript{tr}}}\left(N_{j,t}\times tr\textsubscript{u,j}\right)}{\sum_{i\epsilon{T\textsubscript{u}}} tr \textsubscript{u,i}}
\end{equation}
$E_u$= weighted 300 dimensional embedding for user $u$,\\
$ts_{u,t}$ = the score for tag $t$ for user $u$, \\
$ft_t$ = 300 dimensional fasttext embedding for tag $t$\\
$N_{j,t}$, $T_u$, $V_{tg}$, $V_{tr}$, $tr_{u,i}$ are defined the same as in \eqref{tagscore} and \eqref{n_tagweight}.

Subsequently, Logistic regression with Lasso (or L1) regularisation was used to reduce the feature dimensionality to avoid over-fitting. An SVM(Support Vector Machine) model was then trained to predict the group label(No-risk or At-risk) given the user-feature.

\begin{table*}
 \centering
 \scalebox{0.9}{
 \begin{tabular}{|p{2.1cm}|p{1cm}|p{4.2cm}|p{3cm}|}
  \hline
  Emotion Term&Factor&Emotion Category&GEMS 3-factor label\\
  &Loading&(VAD/VA values)&\\
  \hline
  Happy&1.0&&\\\cline{1-2}
  Amazed&0.95&Wonder&\\\cline{1-2}
  Dazzled&0.84&(6.43, 4.77, 5.92 / 6.29, 4.57)&\\\cline{1-2}
  Allured&0.86&&\\\cline{1-2}
  Moved&0.75&&\\
  \cline{1-3}
  Inspired&1.0&&\\\cline{1-2}
  Transcendence&0.92&Transcendence&\\\cline{1-2}
  Spirituality&0.90&(6.36, 4.70, 5.94 / 6.31, 4.73)&\\\cline{1-2}
  Thrills&0.65&&Sublimity\\
  \cline{1-3}
  In Love&1.0&&\\\cline{1-2}
  Affectionate&0.97&Tenderness&\\\cline{1-2}
  Sensual&0.98&(6.65, 4.62, 6.11 / 6.56, 4.62)&\\\cline{1-2}
  Tender&0.97&&\\\cline{1-2}
  Softened-up&0.74&&\\
  \cline{1-3}
  Sentimental&1.0&&\\\cline{1-2}
  Dreamy&0.77&Nostalgia&\\\cline{1-2}
  Nostalgic&0.64&(5.97, 4.15, 5.57 / 5.92, 3.97)&\\\cline{1-2}
  Melancholic&0.52&&\\
  \cline{1-3}
  Calm&1.0&&\\\cline{1-2}
  Relaxed&0.96&Peacefulness&\\\cline{1-2}
  Serene&0.94&(6.72, 3.1, 6.4 / 6.63, 2.95)&\\\cline{1-2}
  Soothed&0.90&&\\\cline{1-2}
  Meditative&0.58&&\\
  \hline
  Energetic&1.0&&\\\cline{1-2}
  Triumphant&0.76&Power&\\\cline{1-2}
  Fiery&0.72&(6.3, 5.16, 6.12 / 6.39, 5.22)&\\\cline{1-2}
  Strong&0.70&&\\\cline{1-2}
  Heroic&0.56&&Vitality\\
  \cline{1-3}
  Stimulated&1.0&&\\\cline{1-2}
  Joyful&0.99&Joyful Activation&\\\cline{1-2}
  Animated&0.95&(6.8, 5.31, 6.22 / 6.67, 5.43)&\\\cline{1-2}
  Dancing&0.72&&\\\cline{1-2}
  Amused&0.56&&\\
  \hline
  Agitated&1.0&&\\\cline{1-2}
  Nervous&0.85&Tension&\\\cline{1-2}
  Tense&0.63&(3.31, 5.17, 4.0 / 3.38, 5.24)&\\\cline{1-2}
  Impatient&0.49&&Unease\\\cline{1-2}
  Irritated&0.39&&\\
  \cline{1-3}
  Sad&1.0&Sadness&\\\cline{1-2}
  Sorrowful&0.82&(2.99, 4.19, 3.89 / 2.81, 3.61)&\\
  \hline
  
 \end{tabular}}
 \caption{Factor Loadings for the first-order Musical Emotion Categories\cite{gems}. These 9 factors are the emotion categories used in the paper. The VA/VAD values for the 40 emotion terms as presented were obtained from the Tag Emotion Induction model. These were then weighted and summed according to their respective factor loadings to finally obtain VA/VAD values for each of the 9 emotion categories. The 9 factors are further categorised into 3 umbrella categories which are displayed in the fourth column.}
 \label{gemsloading}
\end{table*}

\begin{table*}
 \centering
\resizebox{\textwidth}{!}{%
\begin{tabular}{|l|l|l|l|l|l|}
  \hline
  Group&Top&\multicolumn{2}{|c|}{VAD}&\multicolumn{2}{|c|}{VA}\\\cline{3-6}&Tracks&\textit{t}=$\pm3$&\textit{t}=$\pm2$&\textit{t}=$\pm3$&\textit{t}=$\pm2$\\
  \hline
  &\textit{n}=100&Sadness(11466.0,0.011)&&Sadness(11414.5,0.009)&\\\cline{2-6}
  &\textit{n}=200&Sadness(11672.0,0.018)&Sadness(11928.0,0.043), &Sadness(11525.0,0.013)&Sadness(11881.0,0.037),\\
  At-Risk&&&Tenderness(11873.0,0.037)&&Transcendence(11982.0,0.05)\\\cline{2-6}
  &\textit{n}=500&Sadness(11668.0,0.021), &Tenderness(11876.0,0.034)&Sadness(11617.0,0.016)&Sadness(11958.0,0.049),\\
  &&Tenderness(11905.0,0.04)&&&Transcendence(11962.0,0.049)\\
  \hline
  &\textit{n}=100&&Transcendence(15871.5,0.011)&&Wonder(16270.0,0.003)\\\cline{2-6}
  No-Risk&\textit{n}=200&Transcendence(15698.0,0.022)&Transcendence(15941.5,0.011)&Wonder(15784.5,0.018)&Wonder(16185.0,0.005)\\\cline{2-6}
  &\textit{n}=500&Transcendence(15445.0,0.046)&Transcendence(15879.0,0.015)&Wonder(15738.0,0.02)&Wonder(16139.0,0.004)\\
  \hline
 \end{tabular}
 }
 \caption{MWU test results (U-Statistic, Bootstrap p-value) depicting emotion categories with Significant differences between At-Risk and No-Risk groups.}
 \label{results}
\end{table*}

\begin{table*}
\centering
\resizebox{\textwidth}{!}{%
\begin{tabular}{|l|l|l|}
\hline
\textbf{Cluster} & \textbf{Tags}  & \textbf{Cluster Label} \\ 
\textbf{Number}&&\\\hline

1                                                                                       & 'Extreme Metal', 'experimental metal', 'Extreme', 'true metal', 'Avant Garde Metal'                                                                                              & Avant Garde Metal                           \\ \hline
2                                                                                       & 'House', 'techno', 'club', 'tech house', 'Progressive House', 'electro house'                                                                                                    & Techno/House                                \\ \hline
3                                                                                       & 'electropunk', 'electronic body music', 'belgian', 'electro industrial', 'industrial dance'                                                                                      & Electronic Rock                             \\ \hline
4                                                                                       & 'pop', 'alternative', 'rock', 'indie', 'dance', 'female vocalists', 'alternative rock', 'pop rock'                                                                               & Indie/Alternative Pop-Rock                  \\ \hline
5                                                                                       & '90s dance', 'classic house', 'flash house', 'old school rave', 'eurodance', '90s Eurodance'                                                                                     & 90s Pop/Dance                               \\ \hline
6                                                                                       & 'Swing Jazz', 'lindy hop', 'classic jazz', 'Early Jazz', 'jazz Big Band'                                                                                                         & Swing, BigBand Jazz                         \\ \hline
7                                                                                       & 'jazz standard', 'jazz guitar', 'favourite jazz', 'mellow jazz', 'Blue Note Records'                                                                                             & Chillout-,Easy listening Jazz               \\ \hline
8                                                       & 'Avant Garde Black Metal', 'swedish black metal', 'progressive black metal',& Black Metal \\ &
'satanic black metal', 'Orthodox Black Metal'          &     \\ \hline
9                                                                                       & '80s funk', 'classic uk soul', 'post disco', 'electro disco', '80s soul'                                                                                                         & 80s Soul Funk                               \\ \hline
10                                                                                      & 'medieval folk', 'celtic folk', 'pagan folk', 'neomedieval', 'dream folk'                                                                                                        & Fantasy/Medieval Folk                       \\ \hline
11                                                                                      & 'dream pop', 'noise', 'noise pop', 'shoegaze', 'noise rock', 'noisecore', 'dreampop'                                                                                             & Neo-,Psychedelic-,Dream-Pop                 \\ \hline
12                                                                                      & 'la punk', '80s punk', 'melodic punk', '80s hardcore', 'Horrorpunk', 'psychobilly', 'horror punk'                                                                                & Punk                                        \\ \hline
13                                                                                      & 'finnish', 'finnish metal', 'Sonata Arctica', 'Rhapsody Of Fire', 'medieval metal', 'Power ballad'                                                                               & Symphonic Melodic Metal                     \\ \hline
14                                                                                      & 'world', 'world fusion', 'World Music', 'african', 'world beat', 'mestizo', 'flamenco fusion', 'afrobeat'                                                                        & World Music                                 \\ \hline
15                                                                                      & 'Progressive rock', 'Progressive', 'Psychedelic Rock', 'art rock', 'psychedelic', 'prog rock'                                                                                    & Neo-Progressive Rock                        \\ \hline
16                                                                                      & 'melodic trance', 'uplifting trance', 'progressive trance', 'vocal trance', 'trance-love it'                                                                                     & Chillout Trance                             \\ \hline
17  & 'classic rock', '60s', 'oldies', 'blues rock', 'blues', 'Rock and Roll', '70s', 'rock n roll',&Country and Rock\\ &'americana', 'country', 'country rock'&\\ \hline
\end{tabular}%
}
\caption{Tags along with the corresponding cluster labels. The tags displayed are the ones with high average similarity among the core tags.}
\label{clusters}
\end{table*}

\bibliography{ISMIRtemplate}